\documentstyle[aps,prd,twocolumn]{revtex}

\begin{document}

\title{ Quantization of the Particle Motion 
 on the $n$-Dimensional Sphere}

\author{Petre Di\c t\u a }

\address{Institute of Atomic Physics, P O Box MG6, Bucharest, Romania\\{\rm
email: dita@theor1.ifa.ro}}

\maketitle

\begin{abstract}
We develop here a simple formalism that converts the second-class constraints
into first-class ones for a particle moving on the $n$-dimensional sphere.
The Poisson algebra generated by the Hamiltonian and the constraints closes
and by quantization transforms into a Lie algebra. The observable of the theory is given
by the Casimir operator of this algebra and coincides with the square of the
angular momentum.
\end{abstract}
\section{Introduction} The quantization of classical Hamiltonians, when the
canonical coordinates 
are not completely independent, is a long-standing problem in quantum mechanics.
The constraints, i.e. a set of functions
$$\varphi_i(q,p)=0~~~,~~~ i=1,2,\dots,p \eqno(1.1)$$
restrict the motion of the classical system to a manifold embedded in the initial 
Euclidean phase space. This has in consequence that the canonical quantization rules
$$[q_i,p_j]=i\hbar\delta_{ij}$$
are no more sufficient for the quantum description of the physical system.

When the manifold is a proper subspace of an Euclidean space Podolski
 \cite{Po} gave
a solution by postulating that the Euclidean Laplacian should be replaced by the
Laplace-Beltrami operator acting on this manifold.  Applied to the  motion of a point 
particle on a $n$-dimensional sphere $S^n$ of radius $R$
this gives for the Laplace-Beltrami operator
the result $L^2/2R^2$, where $L$ is the angular momentum
of the particle.

The eigenvalues of the Casimir operator $L^2$ are $l(l+n-1)$,~~ $l=0,1,\dots$, and
in deriving this result one makes use of the Lie algebra of the angular momentum
forgetting completly about the canonical variables initially  entering  in the 
problem, avoiding in this way any trouble which could appear.

Doubts concerning the correctness of this spectrum have been rised by people
who derive the Schr\"odinger equation by Feynman's path integral method; see for example
\cite{Wi,Ch,De}. They found an extra energy term proportional to the Riemann scalar curvature
of the manifold, but they do not agree upon the value of the proportionality factor.

Another type of doubt has been rised recently \cite{Kl-Sb}, namely that the Dirac's quantization
method \cite{Di} has to be rejected because, at least for this problem, the resulting energy spectrum
is physically incorrect.

The last statement \cite{Kl-Sb} comes from a misunderstanding of the sublety of the 
problem: the constraints, like (1), reduce the dimension of the original phase space.
The  mechanism found by Dirac was the introduction of 
a new symplectic structure to handle the second-class
constraints. The first-class constraints are used
 to dropping of some pairs of dinamical variables
$(q_i,p_i)$ from the naive Hamiltonian, whose effect is that the nonphysical 
degrees of freedom are eliminated. See ref. \cite{Fa-Hi} for a treatment of the 
rigid rotator in the Dirac-bracket quantization formalism.

The purpose of this paper is to look at the Dirac formalism from a slightly
modified point of view and to show  that the new proposal leads to correct results.
In fact we propose a new method for converting the second-class constraints into
first-class constraints.

The Dirac quantized theory \cite{Di} is patterned after the corresponding classical
theory: the observables representing constraints  {\it must} have  zero expectation
values. This requirement is inconsistent with the fact  that there are dynamical
variables whose Poisson brackets with the constraints fail to vanish. To solve
the problem Dirac has constructed a new type of bracket, the  Dirac-bracket,
which vanishes whenever one of the two factors is a second-class constraint. 

We develop here a formalism that converts the second-class constraints into
first-class ones and which
leads directly to group properties of the
Poisson brackets.
 By  quantization the Poisson algebra
goes into a Lie algebra. The {\it observables} of the theory will be the
{\it Casimir operators}  of this algebra and the operators generated by the constraints
will commute with the observables in the whole Hilbert space.

With this interpretation the
Dirac's theory of constrained systems gives correct results, and in the particular
case of a point particle on $S^n$ it confirms the Podolski result.

Our idea is to separate all the constants terms which may appear on the left
side of Eq. (1.1) and push them on the right side. Thus we prefer to write (1.1)
in the new form
$$\varphi_i(q,p)=a_i,\quad i=1,2,\dots,p\eqno(1.1')$$
where $a_i$ are some complex constants.

A reason is that the Poisson bracket structure does not discriminate
between $\varphi$ and $\varphi+C$, with $C$ a constant. The main reason is that
it now becomes possible to write the Poisson algebra in a closed form 
like
$$\{\varphi_i(q,p),\varphi_j(q,p)\}=C_{ij}^k\varphi_k(q,p)$$
$$\{H(q,p),\varphi_i(q,p)\}=C_i^j\varphi_j(q,p)\eqno(1.2)$$
where $H(q,p)$  is the Hamiltonian,
$\varphi_i(q,p)$ are the constraints appearing in (1.1')
 and $C_{ij}^k$ and $C_i^j$ are {\it constant} structure coefficients.

Almost all that is found in the Dirac book \cite{Di}, the  novelty
being only the redefinition of the dynamical variables $\varphi_i$.
But this new form
has the advantage of transforming  at least a part of second-class
 constraints into
first-class constraints as we will show in the next section.

In our opinion the Poisson algebra (1.2) generated by the Hamiltonian
and the constraints {\it  is the  best of the Dirac method}.

The Poisson algebra, by the quantization procedure, transforms into
a Lie algebra. The true observables of the physical
systems are given by the Casimir operators of the corresponding Lie algebras.
 In other words no one
of the initial operators does transform into a veritable observable. 
The observables are given, at least, by quadratic functions of the old
operators: Hamiltonian plus constraints together.

The quantic description of the physical model is given by a representation 
of this algebra
onto a Hilbert space. In this way it becomes possible to avoid the canonical
variables, which appear also in the Dirac formalism and, sometimes, cause
problems since their Dirac brackets are not always   canonical; in this sense 
see the treatment of
the three-dimensional  rotator in ref.\cite{Fa-Hi}.

An other consequence of the above idea is a solution of the embarassing situation
of forcing the operators generated by the constraintes to vanish on the whole Hilbert
space, as Dirac dixit. Because  now the constraints are no more observables
the above problem disappers.
What we can say is that there exists a representation of the Lie algebra into an operator algebra acting on
the Hilbert space of the associated physical system such that the operators
$\widehat{\varphi_i}$, generated by the constraints,
should have the numbers $a_i$ in their spectra.

Of course there are cases when the above procedure does not work. An example
of such a constraint is
$$\chi_1 = \sum_{i}c_i\,q_i=a$$
which is linear in coordinates. Its Poisson bracket with
a quadratic free Hamiltonian gives the secondary constraint
$$\chi_2=\{\chi_1,H\}=\sum_i c_ip_i$$
which is linear in momenta. The Poisson bracket of these two constraints
is a constant $\sum c_i^2$, hence the linear constraints in coordinates
and/or momenta generate another type of algebras. In these cases the relations
(1.2) are suplemented with, at least. a few of the form
$$\{\chi_i,\chi_j\}=a_{ij}$$
where $a_{ij}$ are constants. These cases has to be treated by the Dirac formalism.

With this mild interpretation of Dirac theory the difficulties are overcame
and the new theory is ready for applications.

In Sec.2 we treat the point particle on the $n$-dimensional sphere showing
that the "Hamiltonian" of the problem is the square of the angular momentum
and is obtained as the Casimir operator of the  Lie algebra obtained by
quantization of the Poisson algebra generated by the Hamiltonian and constraints.
 In Sec. 3 we consider a related problem: a $(n+1)$-dimensional harmonic oscillator
constrained to move on a hypersurface. The paper ends with Conclusion.

\section{Point Particle on $S^n$ sphere} 
We shall consider
a point particle moving on the $n$-dimensional sphere $S^n$ whose equation is
$$r^2=(q,q)=\sum_{i=1}^{n+1} q_i^2$$
where $(q,q)$ denotes the Euclidean scalar product in the $n+1$-dimensional
space, i.e. we view the particle moving into a subspace of the
$n+1$-dimensional Euclidean space. 
Phase space degrees of freedom $(q_i,p_i)$ take values over the entire real axis
and possess canonical Poisson bracket structure.

The primary constraint is usually written as
$$\varphi=r^2-R^2=0$$
and the Hamiltonian has the form
$$H={1\over 2}(p,p)= {p^2\over 2}$$

We define $U=r^2$ as new dynamical variable  and by taking the Poisson brackets
 we get
$$\{U,H\}= 2\sum q_ip_i= 2 (q,p)= 2V$$
$$\{V,H\}=p^2= 2 H\eqno(2.1)$$
$$\{U,V\}=2 U$$

If we use $\varphi$ as dynamical variable instead of $U$, as it is usually done
\cite{Kl-Sb,Fa-Hi},
we find
$$\{\varphi,H\}=2V$$
$$\{V,H\}=2H$$
$$\{ \varphi,V \}=2r^2=2(\varphi+R^2)$$

The last relation is usually written as $\{\varphi,V\}=2r^2=2R^2$
which has  consequence that after quantization one gets the commutation relation
$[\varphi,V]=2i\hbar\,R^2$. Here we use the same notation for operators on Hilbert
space and for dynamical variables on phase space. The last relation is in conflict
with the conditions $\varphi=V=0$ imposed to  operators acting on Hilbert space.
How we will solve this conflict will be seen later.

The relation $\{\varphi,V\}=2(\varphi+R^2)$ suggested us the introduction
of $U$ as dynamical variable because
$$\{\varphi,V\}=\{\varphi+R^2,V\}=2(\varphi+R^2)$$

Taking $U=\varphi+R^2=r^2$  as dynamical variable has the advantage of closing the algebra as the
relations (2.1) show and, more important, by this procedure both 
$U$ and $V$ become first-class constraints.

 In the sequel we shall take $\hbar=1$.

By the
correspondence principle we obtain from Eqs. (2.1) the commutation relations
$$[U,H]=2 i V$$
$$[V,H]= 2 i H\eqno(2.2)$$
$$[U,V]=2 i U$$
In order to solve the problem we state our first postulate: {\it all the
relevant physics concerning the problem is contained in the Lie algebra (2.2).}

This algebra remind us the known Lie algebra of the $SU(2)$ group, so we can
proceed like in that case. The single observable is the Casimir operator
which is easily seen to be
$$C=V^2-UH-HU \eqno(2.3)$$
Indeed by a trivial calculation we obtain that
$$[C,H]=[C,U]=[C,V]=0$$
$C$ is the true observable of the theory and it commutes with the operators
generated by the classical constraints and the classical Hamiltonian. 

Like in the $SU(2)$ case we can look for a common basis of eigenvectors for
$C$ and one of the operators $H, U$, or $V$ . We choose $V$ as the "third"
component because it is singlet out by the algebra (2.2) as we shall see later.

From Eq.(2.3), $V=(q,p)$ is the scalar product of $q$ and $p$ giving the
projection of the momentum along the radius. The classical requirement
$V=0$ means that the motion of the point particle has to be such that 
there should be no energy or momentum flow across the sphere surface.

If we change from Euclidean to spherical coordinates $V$ has the form
$$ V={1\over i}\,r\,{\partial\over\partial r}$$
where $\partial/\partial r$ is the normal derivative or the gradient at
the point of the sphere determined by its spherical coordinates.

Usually one imposes $V=0$ as an operator equation on the Hilbert space. 
In our approach $V$ is not an observable so the equation $V=0$
is senseless.Our point
of view is expressed by the second postulate: {\it in the Hilbert space
associated to our physical problem there does exist  one vector which is annihilated
by $V$}.

Thus the eigenvector of $C$ and $V$ has to satisfy the equation
$$r\,{\partial\Psi(r,\Omega)\over\partial r}|_{r=R}=0\eqno(2.4)$$
where by $\Omega$ we denote the angular variables. The last equation
tell us that
the eigenvector $\Psi$ may depend on the radial variable $r$, but its
dependence is such that $\Psi$ is stationary at $r=R$. If we require that
Eq.(2.4) should be valid for an arbitrary value of $R$ we get that $\Psi$
does not depend on $r$.

By using the commutation relations (2.2) we find that
$$C=V^2+2iV-2UH$$
On the other side, in spherical coordinates, $H$ has the form
$$2H=-({\partial^2\over\partial r^2} + {n\over r}\,{\partial\over\partial r}
+{1\over{ r^2}} L^2)$$
where $L^2$ is the Laplace-Beltrami operator on the unit Sphere $S^n$ and coincides
with the Casimir operator of the orthogonal group $SO(n+1)$ \cite{Vi}.

By putting together the previous information we find that $C$ has the form
$$
C={L^2 +(n+1)\,r\,{\partial\over\partial r}}\eqno(2.5) $$

Taking into account  Eq.(2.4) we find that
$$C\Psi=(L^2+(n+1)\,r\,{\partial\over\partial r})\Psi|_{r=R} = L^2\Psi=l(l+n-1)\Psi
$$
Although $C$ is not a Hermitean operator, because $V$ is not, 
the action of both $C$ and $V$ on
the eigenvector $\Psi$ reduces to the action of the Hermitean operator $L^2$.
In this way the eigenvalue problem is quantum mechanically well posed and
the spectrum is $l(l+n-1)$, $l=0,1,\dots$, \cite{Vi}.

Thus the Dirac formalism in the new interpretation tell us that the observable of
a  particle moving on the $n$-dimensional sphere is its angular momentum, a result
which everybody expected to be so.
This result can be tested  directly at the classical and by way of consequence
 at quantic level.

For simplicity we take $n=2$ and in this case the components of the angular momentum
are
$$L_1=q_2p_3-q_3p_2,~~~L_2=q_3p_1-q_1p_3,~~~L_3=q_1p_2-q_2p_1$$
Let us introduce its projection on an arbitrary ray
$$S=q_1L_1+q_2L_2+q_3L_3=l\,R$$
where $l$ is a constant, the length of the projection, and consider $S$ as a new
constraint.

Taking into account the Poisson bracket relations
$$\{L_i,p_j\}=\epsilon_{ijk}p_k,~~~~\{L_i,q_j\}=\epsilon_{ijk}q_k$$
$$\{L_i,L_j\}=\epsilon_{ilk}L_k,~~~~\{x_i,p_j\}=\delta_{ij},~~i,j,k=1,2,3$$
we find that
$$\{S,H\}=\{S,U\}=\{S,V\}=0$$
relation which shows that $S$ is a conserved quantity at the classical level
since it commutes with the Hamiltonian and both constraints! At  quantic level
this relation says that $S=f(C)$, i.e. it is a function of the Casimir operator
of the Lie algebra.

We think that the above correct quantization of this simplest non-Euclidean
system will have a fundamental theoretical interest showing us the route to
follow in much more complicated cases.

Now we want to show that $V$ is the analog of $L_3$ of the $SU(2)$ group.
We make the notation
$$H_1=-{V\over 2i}~~~~E_+=-{U\over 2i}~~~~E_{\_}={H\over 2i}$$
and the commutation relations (2.2) take the form
$$[H_1,E_+]=E_+$$
$$[H_1,E_{\_}]=-E_{\_} \eqno(2.6)$$
$$[E_+,E_{\_}]=H_1$$
i.e. the well known Cartan-Chevalley form of the most simple Lie algebra.
From (2.6) we see that $E_{\pm}$ are the analog of $L_{\pm}$ and
$H_1$ is the analog of $L_3$, where $L_{\pm}$ and $L_3$ are the usual
generators
of the $SU(2)$  group.

\section{Constrained Harmonic Oscillator}
In the  
following we give a new argument in  favour of our interpretation. For this
we will consider another simple system: namely
the quantization of the $\ell$-dimensional oscillator whose Hamiltonian is
$$H_0={1\over 2}\sum_{i=1}^{\ell}(p_i^2+q_i^2)$$
We suppose that its movement is confined to the hypersurface given by the
constraint
$$V_0=\sum_{i=1}^{\ell}q_i\,p_i=a\eqno(3.1)$$
where $a$ is a constant.

If we proceed as above we find the Poisson algebra
$$\{H_0,V_0\}=\sum_{i=1}^{\ell}(q_i^2-p_i^2)=2U_0$$
$$\{U_0,V_0\}=2H_0$$
$$\{U_0,H_0\}=2V_0$$
After quantization we find the Casimir operator which commutes with $H_0,
U_0, V_0$

$$C_0=H_0^2-V_0^2-U_0^2$$
We denote by $L_{ij}=q_ip_j-q_jp_i$, $i<j$, $i,j=1,2,\dots,\ell$
the components of the angular
momentum and by using the above expressions for $H_0$,
$U_0$ and $V_0$ we find that
$$C=\sum_{i<j}L_{ij}^2 =L^2$$
i.e. the true Hamiltonian of the problem is again the square of the
angular momentum.
In fact this problem is a reformulation of the previous one. Indeed $H_0,
U_0$ and $V_0$ are related to $H,U,V$ by the relations
$$H_0={1\over 2}U+H$$
$$U_0={1\over 2}U-H $$
$$V_0=V$$

Our procedure can be applied whenever the Poisson algebra closes after
a finite number of secondary constraints. This
may not be the usual situation as a "small peturbation" of the previous
problem shows.
We deform the constraint (3.1) to
$${\cal V}_1=\sum_{i=1}^{i=\ell}c_iq_ip_i=a$$
where $c_i$ are constants.
Then one easily finds that the Poison algebra never closes.
If we  define the sequence of "Hamiltonians"
$${\cal H}_n={1\over 2}\sum_{i=1}^{i=\ell}c_i^{2n}(q_i^2+p_i^2)~~~~~n=0,1,\dots$$
and constraints
$${\cal V}_n=\sum_i c_i^{2n-1}q_ip_i=0\quad n=2,3,\dots$$

$${\cal U}_n={1\over 2}\sum_i c_i^{2n-1}(q_i^2-p_i^2)=0\quad n=1,2,\dots$$
after quantization we find an infinite-dimensional algebra. Its commutation
relations are

$$[{\cal U}_m ,{\cal H}_n]=2i\,{\cal V}_{m+n},~~m=1,2,\dots~~n=0,1,\dots$$
$$[{\cal H}_m,{\cal V}_n]=2i\,{\cal U}_{m+n},~~m=0,1,\dots~~n=1,2,\dots$$
$$[{\cal U}_m,{\cal V}_n]=2i\,{\cal H}_{m+n-1}~~~m,n=1,2,\dots\eqno(3.2)$$
$$[{\cal H}_m,{\cal H}_n]=0$$
$$[{\cal U}_m,{\cal U}_n]=0$$
$$[{\cal V}_m,{\cal V}_n]=0$$
where in the last three equations the indices of ${\cal H}_{m}$ take the values
$m=0,1,2,\dots$, the range of the others being $m=1,2,\dots$, according with the
notation of the first three equations.

A similar algebra is obtained by deforming the sphere $(q,q)=R^2$ into an 
ellipsoid by the change $q_i\rightarrow{q_i/a_i}$. This shows that the problem
of quantization with constraints is not a simple one,  its natural 
place being the  representation theory of infinite-dimensional algebras.
\section{Conclusion}
The main difficulty appearing in quantization of constrained systems is caused by 
second-class constraints. To overcome it Dirac invented a new symplectic
structure, the Dirac bracket. However its use is not straightforward and
we have to take care of when using it. This was a sufficient reason to looking
 for new methods of quantization.

The best known one is the method of abelian conversion that transforms 
a second-class constrained system into an abelian gauge theory \cite{Fa-Sh,Ba-Fr}.
The idea is to extend the original phase space by introducing new canonical
coordinates and to convert the original Hamiltonian into a new one obtained by
solving some equations. Upon quantization all non-physical degrees of freedom
are removed  by a restriction of the Hilbert space to a physical subspace formed by
gauge invariant states.

In this paper we observed that there are some cases when the conversion of second-class
constraints into first-class ones is very simple, namely we have seen that
the obstacle was caused by the presence of constant terms within the functions defining
the constraints. Because the Poisson bracket of a dynamical variable with a
constant vanishes this opens the possibility to rewrite the original Poisson 
brackets into a new form by a simple redefinition of some of the dynamical variables.
In this way it becomes possible to write the Poisson brackets into the form
of Poisson algebra (1.2), that, after quantization, transforms into a Lie algebra.
Once obtained this algebra we can use the known powerful machinery of representation
theory to find the {\it observables} of the physical theory formalized
by this algebra and to obtain the spectra of the physically relevant operators.

The lesson to be learned is that for constrained systems no one of the initial
dynamical variables  transforms into an observable. The observables are given
by   Casimir operators of  Lie algebras, i.e. at least by quadratic
functions of dynamical variables. By way of consequence we are no more constrained
to impose operators equations on the Hilbert space, like in our case $\varphi=V=0$,
because the constraints do not become {\it observables} of the quantum theory.

It will be an interesting exercise to find a physical interesting constrained system
for which the Cartan subalgebra of the corresponding Lie algebra is two-dimensional,
because in this case it will be possible  to obtain two observables of the system.

It seems to be a big challenge the construction of an operator representation
of the infinite-dimensional algebra (3.2) on a Hilbert space that will solve, for
example, the quantization of the motion of a point particle on an ellipsoid. 

\end{document}